\documentclass[aps,prl,showpacs,twocolumn,superscriptaddress]{revtex4-1}
\usepackage[fleqn]{amsmath}
\usepackage{graphicx}
\usepackage{natbib}
\usepackage{color}
\usepackage[plain]{fancyref}

\def \lstar{\ell_\ast}
\def \lh{\ell_H}

\def \lp{\ell_\perp}
\def \ll{\ell_L}
\def \lsf{\ell_{sf}}
\def \lup{\ell_\uparrow}
\def \ldown{\ell_\downarrow}
\def \lstar{\ell_\ast}
\def \mus {\boldsymbol \mu}
\def \muc {\mu_c}
\def \tp{\theta}
\def \jc{j_\alpha^c}
\def \js{\boldsymbol{j}_\alpha}
\def \jq {j_\alpha^q}
\def \tv {\tau_{V}}
\def \tt {\tau_{M}}
\def \toc {\tau_{P}}
\def \mag{\boldsymbol m}
\def \da{\partial_\alpha}

\def \L {\mathcal{L}}
\def \Tav {T^*}
\def \k {k_B}

\def \ds {\Delta s}

\def \Rsh{{\cal R_{\rm Sh}}}
\def \mx {{\rm mx}}

\def \dsum {\displaystyle \sum}

\newcommand {\bv}[1]{\boldsymbol {#1}}
\newcommand {\e}[1]{\boldsymbol e_{#1}}
\begin{document}
\title{Sustained RF oscillations from thermally induced spin-transfer torque}
\newcommand{\spsms}{CEA-INAC/UJF Grenoble 1, SPSMS UMR-E 9001, Grenoble F-38054, France}
\author{David Luc}
\affiliation{\spsms}
\author{Xavier Waintal}
\affiliation{\spsms}
\date{\today}

\begin{abstract}
We investigate the angular dependence of the spin torque generated when applying a temperature difference across a spin-valve. Our study shows the presence of a non-trivial fixed point in this angular dependence,
\textit{i.e.} the possibility for a temperature gradient to stabilize radio frequency oscillations without the need for an external magnetic field.
This so called "wavy" behavior can already be found upon applying a voltage difference across a spin-valve but we find that this effect is much more pronounced with a temperature difference.
Our semi-classical theory is parametrized with experimentally measured parameters and allows one to predict the amplitude of the torque with good precision.
Although thermal spin torque is by nature less effective than its voltage counterpart, we find that in certain geometries, temperature differences as low as a few degrees should be sufficient to trigger the switching of the magnetization.%
\end{abstract}
\maketitle

Spin caloritronics\cite{Bauer:2012, Bauer:2007,Jia:2011,Uchida:2008,Uchida:2010,Uchida:2013,Jia:2013} studies the interplay of charge, spin and heat transport and provides extensions to some of the spintronics concepts.
One of interest to us is the spin-transfer torque (STT)\cite{Ralph:2008, Myers:1999, Katine:2000}, first predicted by Slonczewski and Berger in 1996\cite{Slonczewski:1996, Berger:1996}. STT is the  angular momentum deposited by a spin-polarized current on a ferromagnetic layer. It is at the origin of interesting out of equilibrium dynamics for the magnetization layer leading to magnetic reversal or sustained RF oscillations. The later effect, known as spin-torque oscillator (STO)\cite{Zhang:2003, Houssameddine:2009} is a promising candidate for agile RF sources. Although most STO require an external magnetic field, it was also discovered that STT can, in some very asymmetric spin-valves, stabilize an oscillating state in the absence of an external magnetic field. This is the so-called waviness\cite{Boulle:2008, Waintal:2009, Barnas:2009}. In 2007, in one of the first article on "caloritronics", Bauer \textit{et al.} considered another route for creating STT via the combination of spintronics with
thermoelectric effects\cite{Bauer:2007}: the so-called thermal STT.  Spin-dependent thermoelectric effects soon started to attract some theoretical and experimental interest \cite{Slonczewski:2010,Uchida:2008, Uchida:2010, Uchida:2013, vanWees:2012b,vanWees:2012a}

In this letter, we investigate the angular dependence of the STT induced by temperature gradients applied across various type of magnetic spin valves. Our semi-classical theory, carefully tabulated with experimentally measured parameters, shows that thermally-induced STT is naturally "wavy" for a wide range of devices. By optimizing the geometry of the sample, we predict that magnetic switching can be obtained with temperature differences as low as a few degrees.

{\it Semi-classical drift-diffusion approach.} Our starting point is a semi-classical approached for metallic magnetic multilayers that treats the charge degrees of freedom at the drift-diffusion level yet retains all the information about spin degrees of freedom\cite{Waintal:2000,Waintal:2009}. This approach to which we refer as CRMT\cite{Waintal:2000,Waintal:2009,Waintal:2011} (for Continuous Random Matrix Theory) can be seen as a generalization of the Valet Fert theory\cite{Valet:1993} to systems with non collinear magnetization\cite{Petitjean:2012}. It is also equivalent to the so-called (Generalized) Circuit Theory\cite{Bauer:2003}. Here we generalize CRMT to include heat flow and thermoelectric effects. In addition to the charge $I_\alpha$ and spin $\boldsymbol{J}_\alpha$ current densities, we therefore add the heat current density $Q_\alpha$ ($\alpha=x,y,z$ being the direction of propagation).
Similarly, in addition to the charge $\muc$ and spin $\mus$ potentials, we include the temperature $\theta$ (in energy unit, $\tp=\k T$ where $T$ is the actual temperature). Note that in this letter, we assume that a single temperature can be defined for both majority and minority electrons. Thermoelectric effects are described by spin dependent
Seebeck and Peltier coefficients\cite{vanWees:2010b, vanWees:2011, vanWees:2012b, Uchida:2008}. We note $S_{\uparrow}$ ($S_{\downarrow}$) the spin-dependent Seebeck coefficients for majority (minority) electrons while the Peltier coefficients are given by Onsager relation $\Pi_\sigma=S_\sigma \Tav$ where $\Tav$ is the average temperature.
We further introduce dimensionless Seebeck coefficients in unit of $k_B/e\approx 80\mu V.K^{-1}$:
$s=e (S_{\uparrow} + S_{\downarrow})S/(2\k)$ and $\Delta s=e (S_{\uparrow} - S_{\downarrow})S/(2\k)$ characterize respectively the average and the polarization of the Seebeck effect. Recent experiments provide the first spin resolved
values of these quantities for ferromagnetic materials\cite{vanWees:2012b}: $s_{Co}\approx-0.25$ and $\ds_{Co}\approx -0.02$ for cobalt, and $s_{Py}\approx-0.21$ and $\ds_{Py}\approx -0.044$  for permalloy.
We introduce reduced currents (with unit of energy) as follows,
\begin{align}
    I_\alpha &= 4\jc / (e\Rsh) \\
    \boldsymbol{J}_\alpha &= 2\hbar\js / (e^2\Rsh) \\
    Q_\alpha &= 4k_B \Tav \jq/(e^2\Rsh)
\end{align}
where $\Rsh$ is the Sharvin resistance for a unit surface ( with typical value $\Rsh \approx 1 f\Omega.m^2$), and $e<0$ is the charge of the electron. These variables follow a set of Ohm-like (or Fourier-like) equations,
\begin{widetext}
\begin{align}
-\lstar \da \muc &= \jc-\beta \js \cdot \mag + \dfrac{\lstar}{\lh}s\left( s\jc + \ds \js \cdot \mag \right) - \dfrac{\lstar}{\lh}s \jq\label{eq:crmt1}\\[5pt]
-\lstar \da \mus &= \js-\beta \jc \mag + \dfrac{\lstar}{\lh}\ds\left( s\jc \mag + \ds \js \right)- \dfrac{\lstar}{\lh}\ds \jq \mag+ \dfrac{\lstar}{\lp}\left(\mag \times \js \right)\times \mag - \dfrac{\lstar}{\ll}\left(\mag \times \js \right)\label{eq:crmt2}\\[5pt]
-\lh \da \tp &= -s\jc - \ds \js\cdot\mag+\jq\label{eq:crmt3}
\end{align}
\end{widetext}
Eqs(\ref{eq:crmt1}-\ref{eq:crmt3}) are the extension of Eqs.(1)-(4) of \cite{Petitjean:2012}. The unit vector $\mag$ is the local direction of the magnetization (bold vectors correspond to spin space while explicit components
$\alpha=x,y,z$ are used for real space). The parameters involved are the mean free paths for the majority ($\lup$) and minority ($\ldown$) electrons, related to the spin-dependent resistivities $\rho_{\sigma}$ as $\ell_{\uparrow(\downarrow)} = \Rsh/\rho_{\uparrow(\downarrow)}$. They can be expressed alternatively in term of $\lstar$, the average mean free path ($1/\lstar = 1/\lup+1/\ldown$), and $\beta=(\lup-\ldown)/(\lup+\ldown)$, the asymmetry of the spin resolve asymmetry (with a definition identical to the usual Valet-Fert parameter). Two length scales characterize the behavior of a spin perpendicular to the magnetization: the Larmor precession length $\ll$ and the transverse penetration length $\lp$, see \cite{Petitjean:2012}. Finally, $\lh$ is the heat diffusion length. For purely electronic heat transfer Wiedemann-Franz law implies,
$\lh=\lstar( \L-s^2+2\beta s \ds+\ds^2)/(1-\beta^2)$ with $\L=\pi^2/3$. However, to account for the phonon contribution, higher values of $\L$ can be used.
 A second set of equation expresses the conservation (or lack thereof) of the different currents,
\begin{align}
\sum_\alpha \da \jc &= 0\label{eq:crmt4} \\
\sum_\alpha \da \jq &= 0\label{eq:crmt5} \\
\sum_\alpha \da\js &=  -\dfrac{\lstar}{\lsf^2}\mus - \dfrac{1}{\lp}\left(\mag \times \mus \right)\times \mag + \dfrac{1}{\ll}\left(\mag \times \mus\label{eq:crmt6} \right)
\end{align}
where $\lsf$ is the spin diffusion length. Similarly, a set of equations describe the interface boundary conditions
between a ferromagnet and a normal metal. The charge and spin sectors are described by the usual spin dependent
interface resistances $r^b_\sigma$, namely Equation (8) and (9) of Ref.\cite{Petitjean:2012}. The heat sector
is given by (neglecting interface thermoelectric effects),
\begin{align}
    \sum_\alpha n_\alpha \jq &= \L \dfrac{\Rsh}{4r^b_\ast (1-\gamma^2)} (\tp_N - \tp_F)\label{eq:interface_th}
\end{align}
Where $\tp_N$ and $\tp_F$ are the temperatures on both sides of the Ferro-Normal interface and the interface resistances have been parametrized according to the usual Valet-Fert notation
$r^b_{\uparrow,\downarrow} = 2r^b_\ast(1\pm\gamma)$. $n_\alpha$ are the components of the normal unit vector pointing towards the magnetic side of the interface. Last, the boundary conditions at the metallic electrodes are given by Eqs.(12) and (13) of
Ref.\cite{Petitjean:2012} for the spin and charge sector while the heat sector reads ($n_\alpha$ points towards the system),
\begin{equation}
    \sum_\alpha n_\alpha \jq + \tp = k_B \Delta T \label{eq:boundary}
\end{equation}
where $k_B \Delta T$ is the temperature difference applied to the reservoir with respect to the reference temperature $T^*$.

{\it Application to thermally induced STT in a spin-valve} Let us now turn to a spin-valve made of the following stack:
$\rm{Cu_{20}|Co_{L_{Co}}|Cu_{2}|Py_{L_{Py}}(\varphi)|Cu_{10}}$ where the indices indicate the corresponding thicknesses in nm and $\varphi$ is the angle of the magnetization of the free (permalloy) layer with respect to the fixed cobalt layer.  Following usual practice\cite{Slonczewski:1996, Waintal:2000}, the torque $\bv\tau$ exerted on the free layer is defined as the difference of spin currents on both side of the layer (spin relaxation only provides extremely small corrections here, see\cite{Petitjean:2012}). We used standard material parameters for the mean free paths and spin-diffusion lengths of Cu, Co and Py, as extracted from giant magneto resistance measurements\cite{Petitjean:2012})
while we focus on the values given in Ref\cite{vanWees:2012b} for the spin resolved Seebeck coefficients (see supplementary material).
\begin{figure}[!h]
    \includegraphics[width=\linewidth]{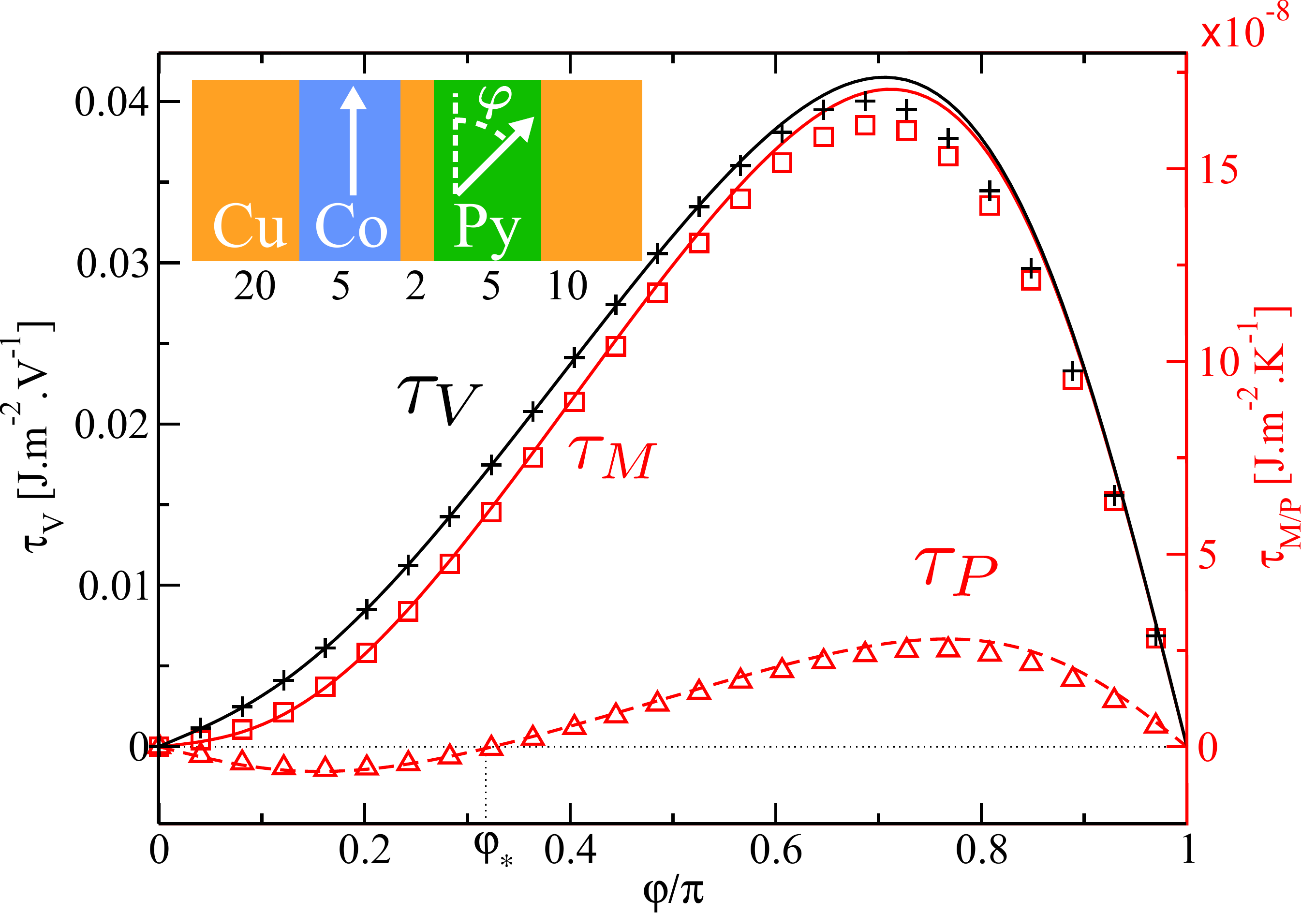}
    \caption{Spin-transfer torque obtained when applying a voltage ($\tv$, bottom curve), a temperature gradient ($\tt$, top full curve), and a temperature gradient in the open-circuit configuration ($\toc$, top dashed curve), versus the magnetization angle $\varphi$ of the Py layer with respect to that of the Co layer. Symbols represent the simulations including spin-flip scattering, while lines correspond to the analytical calculation \Fref{eq:torque}. Here $L_{Co}=L_{Py}=5$ nm. Inset: sketch of the spin valve.
    }
    \label{fig:torques}
\end{figure}

\begin{figure}[!h]
    \includegraphics[width=\linewidth]{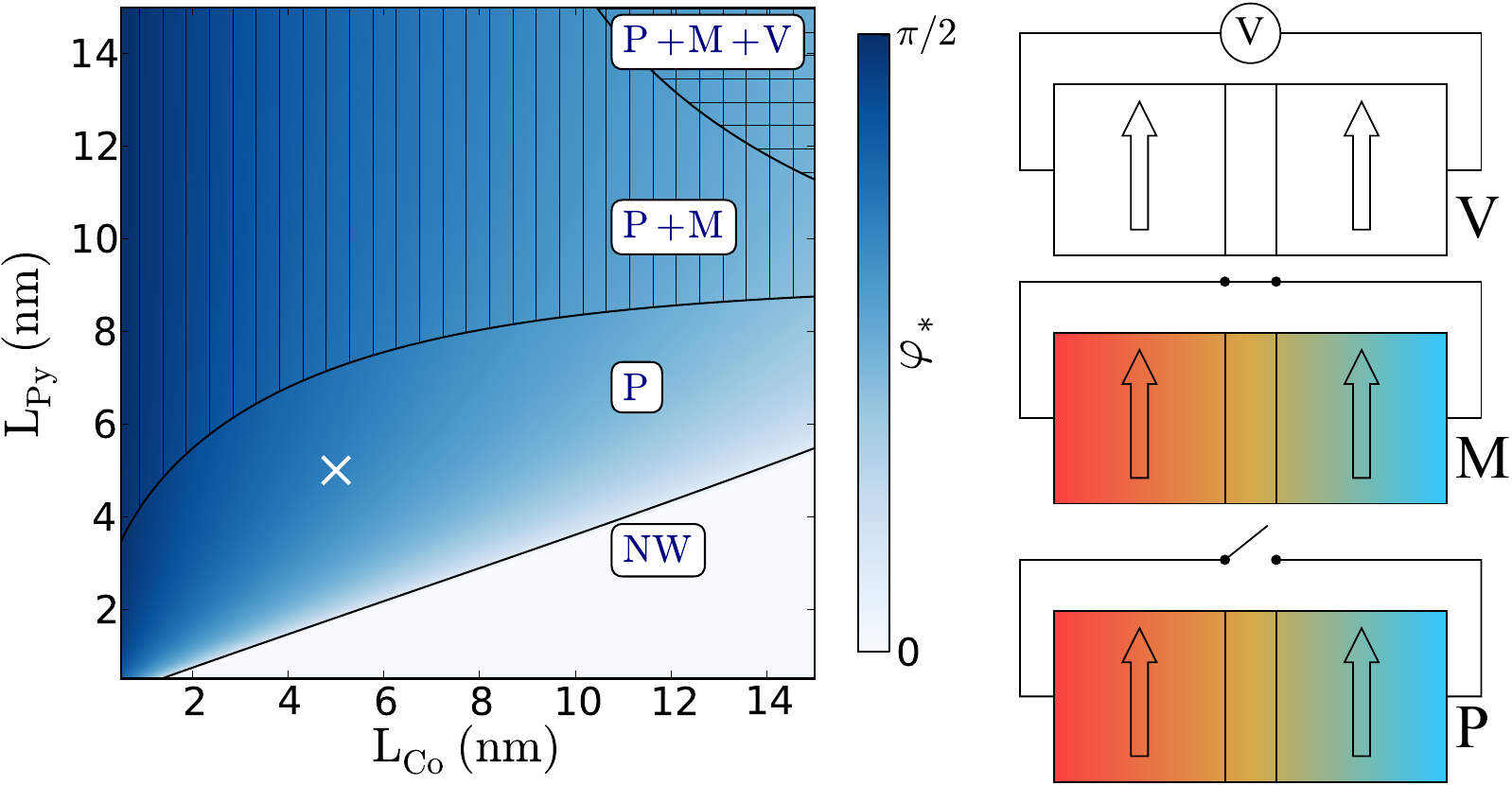}
    \caption{Left: Waviness angle $\varphi^*$ of the pure thermal torque $\toc$ as a function of $L_{Co}$ and $L_{Py}$. The white cross indicates value $L_{Co}=L_{Py}=5$ nm corresponding to \Fref{fig:torques}. The presence of a letter V, M or P in a given region means that the angular dependence of the corresponding torkance $\tv$, $\tt$ or $\toc$ is wavy. NW indicates the region where none of them are wavy. Right: cartoon of our three measurement setups V, M, and P. In M and P a temperature difference is applied across the pillar.}
    \label{fig:phase diagram}
\end{figure}

\Fref{fig:torques} shows the angular dependence of the spin torque for three different types of setups (see the right part of Fig.\ref{fig:phase diagram} for a cartoon). In the first, we apply a voltage bias $V_b$ across the spin valve and calculate the torkance $\tv = d\tau/dV_b$. We recover the usual feature of STT in metallic spin valve with a stronger torque in the anti-parallel configuration than in the parallel one (black curves). In the second, we apply a temperature difference $\Delta T$ across the spin valve in an open
circuit configuration so that no current can flow through the device. This is the "pure" spin Seebeck case $\toc =d\tau/d \Delta T$ as it is given by spin current only. In the last closed circuit or "mixed" configuration, a temperature difference is applied and a current can flow through the spin valve (i.e. the two electrodes of the spin valve are electrically - but not thermally - short circuited). In this last configuration the Seebeck effect induces a finite
current density which in turn induces a STT very similar to the voltage driven one. Hence, one find that the mixed
thermal torkance $\tt =d\tau/d \Delta T$ is somehow intermediate between the pure and the voltage torkances.
The most remarkable feature of \Fref{fig:torques} is the appearance in the pure case $\toc$ of a finite angle
$\varphi^* \approx \pi/3$ where $\toc$ vanishes. Depending on the sign of the thermal gradient, this new fixed point
will be stabilized or destabilized. When stabilized, it corresponds to a fast precessional state which forms a STO.
In the context of voltage induced torque, these "wavy" structures, which do not require magnetic field in contrast to
more conventional STOs, have been discussed for highly asymmetric spin valves\cite{Boulle:2008, Waintal:2009, Barnas:2009}. Here we find that thermally induce STT corresponds to a wavy angular dependence of the torque in a much broader range of parameters. \Fref{fig:phase diagram}, shows the "phase diagram" of the spin valve as a function of the thicknesses $L_{Co}$ and $L_{Py}$ of the fixed cobalt and free permalloy layers. The various regions correspond to the presence of a wavy angular dependence of the torque for thermal induced spin torque (P and M) and the standard voltage induced STT (V). The color measures the waviness angle of $\toc$, when it exists.
This diagram illustrates several points, the first of which is that thermally induced torque is wavy in a much broader
range of thicknesses than the voltage induced torque. Second, the various torques behave quite differently. A thicker Co layer is beneficial for the waviness of $\tv$, whereas it is detrimental for that of $\toc$ and $\tt$. Also, for the limit of a very thin Co layer, the waviness angle for $\toc$ comes close to $\pi/2$. As a comparison, the maximum waviness angle in this diagram for $\tv$ (not represented) is five times lower.

To proceed, we introduce a minimum model to estimate the critical value of the temperature gradient needed to trigger magnetic switching or STO behavior. In the macrospin approximation in presence of a purely uniaxial anisotropy, the critical torque (per unit angle and per unit surface of the spin valve) needed to destabilize the initial (parallel or anti-parallel) configuration is given by \cite{Slonczewski:1996,Waintal:2006},
$\partial \tau/\partial \varphi = \alpha M_s L_{Py} B_u $
where $B_u$ is the uniaxial anisotropy field, $\alpha$ the Gilbert damping coefficient and $M_s$ the magnetization.
Using $\tau = \toc \Delta T$, we obtain the critical value of the temperature gradient $\Delta T_P$ needed to get magnetic switching (or STO) as,
\begin{equation}
\Delta T_P = \frac{\alpha M_s B_uL_{Py}}{\partial \toc/\partial \varphi} \approx
\frac{L_{Py}}{\partial \toc/\partial \varphi} \times 1.67\mathrm{\ kJ \cdot m^{-3} \cdot rad^{-1}}
\end{equation}
The numerical value of the right hand side of the previous expression was obtained by simulating the spin-valve $\rm{Py_{24}|Cu_{10}|Py_{6}(\varphi)}$ of \cite{BassPratt:2004} for which a critical switching current $\rm{I_{crit}=10^7 A\cdot cm^{-2}}$ has been reported. We calculate a corresponding critical torque of the order of $10^{-5}\rm{J\cdot m^{-2}\cdot rad^{-1}}$ which allows us to estimate globally the product $\alpha M_s B_u$.
Critical currents of the order of  $10^7 \rm{A\cdot cm^{-2}}$ are rather standard values for current driven STT\cite{Jiang:2004, Zhang:2002, BassPratt:2004} and values up to two orders of magnitude smaller have been reported\cite{Yang:2008}, so that the previous expression is a rather conservative estimate.

\begin{figure}[!h]
    \includegraphics[width=\linewidth]{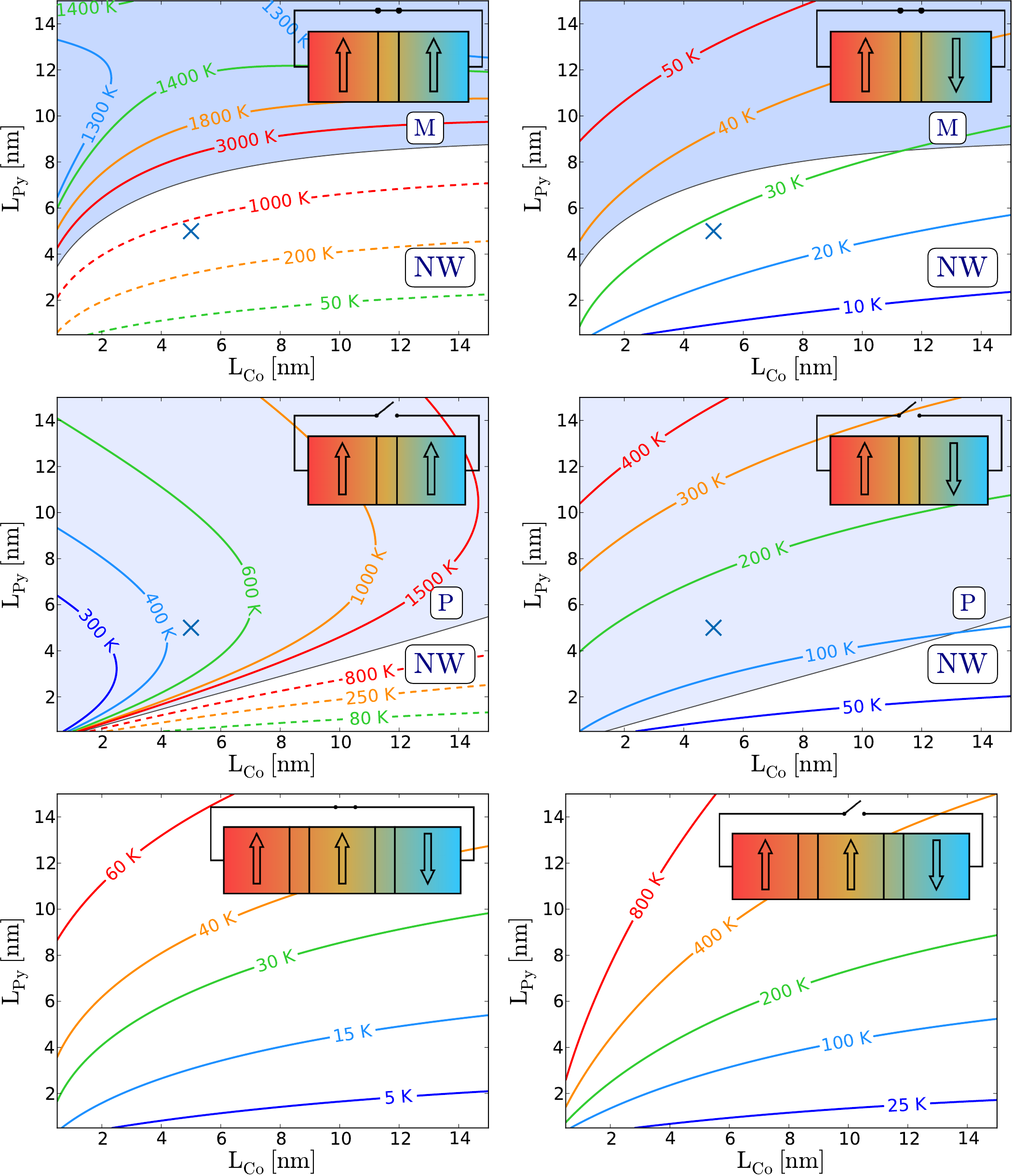}
    \caption{Temperature difference $\Delta T$ needed to achieve the critical torque as a function of $L_{Co}$ and $L_{Py}$ for the mixed ($\Delta T_M$, top row) and pure ($\Delta T_P$, middle row) torques, for a parallel (left column) and antiparallel (right column) configuration, in the $\rm{Cu_{40}|Co_{L_{Co}}|Cu_{2}|Py_{L_{Py}}(\varphi)|Cu_{10}}$ stack. Dashed ligns indicate a negative torkance. The blue cross indicates $L_{Co}=L_{Py}=5$ nm, {\it cf} \Fref{fig:torques}. The background displays the waviness domains of \Fref{fig:phase diagram}. The bottom row shows the same critical temperature difference for the stack $\rm{Cu_{40}|Co_{L_{Co}}|Cu_{2}|Py_{L_{Py}}(\varphi)|Cu_{2}}|Co_{L_{Co}}|Cu_{40}$ for the mixed (left) and pure (right) torques.
    }
    \label{fig:critical temperature}
\end{figure}

\Fref{fig:critical temperature} shows the critical temperature difference in the mixed ($\Delta T_M$, top row) and pure case $\Delta T_P$, middle row). The corresponding values for the parallel configuration (left column) are very high and are not reasonable for actual devices using the reported values for the Seebeck coefficients (although one should bear in mind that careful tuning of the material/geometrical properties could be use to decrease these values significantly). However, the temperature gradient needed to destabilize the anti-parallel configuration (right panel) is much smaller and should be within experimental grasp (a few tens of Kelvin).
To further decrease the critical temperature, we consider a slightly different stack (bottom row of \Fref{fig:critical temperature}) where a third magnetic layer, antiparallel to the first (polarizing) layer has been introduce to enhance the torque. Such an extra layer makes the system perfectly symmetric and therefore makes the waviness behavior disappear. On the other hand, we find a very significant lowering of the critical temperature down to values of a few Kelvin. We find such a low threshold for magnetic reversal to be very encouraging.

{\it Analytical approach: building efficient effective materials}.  In the absence of spin-flip scattering, ignoring the finite penetration of transverse spins and keeping only the first order terms from the Seebeck/Peltier effect, close analytical expressions can be obtain for our model.
A first result is that many collinear materials (or interfaces) put in series can be combined to obtain a unique effective material. After such a procedure our spin valve can be reduced to two effective layers A and B whose magnetizations make an angle $\varphi$. The effective parameters read,
\begin{align}
    &\overline{r} = \dsum_i r_i\label{eq:eff_r}\\
    &\overline{r}\overline{\beta} = \sum_i r_i \beta_i\label{eq:eff_beta}\\
    &\overline{r}(1-\overline{\beta}^2)/{\overline \L} = \dsum_i r_i(1-\beta_i^2)/{\L_i}\label{eq:eff_L}\\
    &{\overline r}\left(1-\overline\beta^2\right) {\overline {s}}/{\overline{\L}}= \dsum_i r_i(1-\beta_i^2) {s_i}/{\L_i}\label{eq:eff_s}\\
    &{\overline r}\left(1-\overline\beta^2\right) {\overline {\ds}}/{\overline{\L}}= \dsum_i r_i(1-\beta_i^2)
    {\ds_i}/{\L_i}\label{eq:eff_ds}
\end{align}
where the resistance $r_i$ of a bulk layer (interface) is given by the ratio $2\rho_\ast L_i/\Rsh$, where $L_i$ is the thickness of the layer ($2r_b^\ast/\Rsh$). These equations can form the basis for engineering the effective parameters and increase the torkance of the stack. By placing two of these effective materials A and B in series, we obtain a  general description of a spin-valve $F_A|N|F_B(\varphi)$. After some algebra, we obtain the expression of the torque on layer B,
\begin{align}
    &\tau = -\dfrac F 2 \sin\varphi\left\lbrace\left[\beta_A\dfrac{r_B+1}{r_B}-\beta_B\cos\varphi \right] \left(GY eV_b+S k_B \Delta T \right)\right.\nonumber\\
    & +\left.\left[\dfrac{\Delta s_{A}}{\L_A}(1-\beta_A^2)\dfrac{r_B+1}{r_B}-\dfrac{\Delta s_{B}}{\L_B}(1-\beta_B^2)\cos\varphi \right]K k_B\Delta T \right\rbrace
    \label{eq:torque}
\end{align}
where $F$, $G$, $Y$, $K$ and $S$ are expressions involving the various material parameters whose sign do not change
when $\varphi$ varies (see supplementary material for explicit expressions). We have checked \Fref{eq:torque} against our numerical simulations and found excellent agreement (see \Fref{fig:torques}).
We find that the current $j^c\propto GY eV_b+S k_B\Delta T$ so that the open-circuit condition for $\toc$ is obtained using $GY eV_b+S k_B\Delta T=0$. While the expression of \Fref{eq:torque} is somewhat cumbersome,
the analysis of its angular dependence allows one to obtain simple criteria for the existence of a wavy regime. We find,
\begin{align}
    \cos\varphi_\ast^V &= \dfrac{\beta_A}{\beta_B}\dfrac{r_B+1}{r_B}\label{eq:wavyV} \\
    \cos\varphi_\ast^{P} &= \dfrac{1-\beta_A^2}{1-\beta_B^2}\dfrac{\L_B}{\L_A} \dfrac{\ds_A}{\ds_B}\dfrac{r_B+1}{r_B}\label{eq:wavyOC}
\end{align}
where the above expressions provide first a criterion for waviness ($|\cos \varphi_\ast| \leq 1$) and second the value
of  $\varphi_\ast$ for wavy structures. We find that the criterion for waviness in the "pure" thermal case contains two conflicting contributions: in order to obtain a wavy structure one needs the polarization of the resistivity of the free (B) layer to be small while the corresponding Seebeck coefficient is highly spin polarized. As both are not necessarily correlated (the former is related to the polarization of the density of state while the latter to its variation with respect to energy), this leaves much room for material optimization.

{\it Conclusion.} We have developed a quantitative theory for spin dependent Seebeck and Peltier effects in magnetic metallic devices. The theory relies entirely on measured material parameters so that its results do not depend on a - always precarious - detailed microscopic modeling. We find that temperature gradient as low as a few degrees should be enough for magnetic switching. Such low temperature gradients could be used in spintronics devices, either alone or to assist current induced switching.

{\it Acknowledgement } Funding was provided by the FP7 project STREP MACALO and the consolidator ERC grant MesoQMC.

\bibliographystyle{apsrev}
\bibliography{../Thermal_torque}
\newpage

\def \mupA {\mu_{\parallel A}}
\def \mupB {\mu_{\parallel B}}
\def \jpll{j_{\parallel}}
\def \jA{j_{\parallel A}}
\def \jB{j_{\parallel B}}
\def \jN{j_{\parallel N}}
\newcommand \dx[1]{\frac{d#1}{dx}}
\def \dsp {\Delta s^\prime}
\def \sp {s^\prime}
\newcommand {\X} [1]{\dfrac{1}{2r_#1}}
\def \angle{\varphi}
\renewcommand{\theequation}{SM-\arabic{equation}}

\section{Derivation of the spin torque expression Fref{eq:torque}}
\Fref{eq:torque} was derived in the case of a spin-valve $F_A| N | F_B(\varphi)$,  under the following assumptions: (i) the spin-valve has no variations along the $y$ and $z$ directions, (ii) spin-flip scattering is neglected,  (iii) the transverse spin is absorbed at the Normal-Ferro interface ($\lp$ very short) , (iv) the Seebeck coefficients $s$ and $\ds$ are only considered at first order and (v) the various layers that make the two effective materials $F_A$ and $F_B$ have a single orientation of their magnetization. The normal spacer is taken to be perfectly transparent without loss of generality as any finite resistance can be incorporated in the effective material $F_A$ or $F_B$. We note that in the numerics presented in the main text, condition (ii) and (iv) are relaxed which only lead to small corrections to the results.

\noindent Within this set of approximations, \Fref{eq:crmt1} to \eqref{eq:crmt6} become for each material:
\begin{align}
-\lstar \dx \muc &= j^c-\beta \jpll - \sp j^q\label{eq:bulk1}\\
-\lstar \dx{\mu_\parallel} &= \jpll-\beta j^c - \dsp j^q\label{eq:bulk2}\\
-\lstar \dx \tp &= -\sp j^c - \dsp \jpll+\frac{(1-\beta^2)}{\L}j^q\label{eq:bulk3}
\end{align}
and the conservation equations are:
\begin{align}
\dx {j^c} &= 0\label{eq:cont1}\\
\dx {j^q} &= 0\label{eq:cont2}\\
\dx\jpll &=  0\label{eq:cont3}
\end{align}
with $\jpll=\bv{j} \cdot \mag$, $\mu_\parallel=\boldsymbol{\mu} \cdot \mag$,  $\sp = \dfrac{1-\beta^2}\L s$ and $\dsp = \dfrac{1-\beta^2}\L \ds$

The conservation equations imply that $j^c$ and $j^q$ are constant, and the absence of spin-flip makes $\jpll$ piecewise constant. As a consequence, $\muc$, $\tp$ and $\mu_\parallel$ are piecewise linear so that \Fref{eq:bulk1} to \Fref{eq:bulk3} can be easily integrated leading to the effective materials described in Eqs.(\ref{eq:eff_r}) -
(\ref{eq:eff_ds}). The matching of spin accumulation of the $a$ ferromagnet with the normal spacer is described by,
\begin{align}
n_x j^c_a &= \dfrac{\Rsh}{4r^b_\ast} \Delta \muc\\
n_x \bv{j}_a &= \dfrac{\Rsh}{4r^b_\ast} \left(\Delta \mus\cdot \mag\right)\mag + \varepsilon_a\left(\mag\times \mus_a\right)\times \mag \label{eq:interface}\\
n_x j^q_a &= \L \dfrac{\Rsh}{4r^b_\ast} \Delta \tp
\end{align}

with $n_x=-1$ ($+1$) for the $F_A|N$ ($N|F_B$) interface, $r^b_\ast$ the square resistance of the interface, $\Delta X = X_N-X_F$ representing the difference of a quantity between the spacer and the ferromagnetic side, and $\varepsilon_N=-\varepsilon_F = 1$. The subscript $a$ indicates on which side of the interface the quantity are evaluated ($N$ or $F$).

Taking the limit of a transparent interface translates to $r^b_\ast \rightarrow 0$. This yields $\Delta \mu_c=0$, $\Delta \mus\cdot\mag=0$ and $\Delta \tp=0$. Applying \Fref{eq:interface} twice, and eliminating all the variables linked to the spacer provides the matching conditions for the spin currents and accumulations between $F_A$ and $F_B$. Specifically, denoting $\mus_A$ (resp $\mus_B$) the vector spin accumulation in layer A (resp. B) infinitely close to its interface with the normal spacer. We have $\mus_A = \mu_A \mag_A = \mu_A \boldsymbol{e}_z$, and $\mus_B = \mu_B \mag_B = \mu_B \left(\sin\varphi \boldsymbol{e}_x + \cos\varphi \boldsymbol{e}_z\right)$. Introducing $\mu_x$ and $\mu_z$ the components of the spin accumulation inside the normal spacer, we get:
\begin{align}
&\mu_z = \mupA \\
&\mu_x\sin\varphi + \mu_z\cos\varphi = \mupB \\
&j_z = \jA \\
&j_x\sin\varphi + j_z\cos\varphi = \jB\\
&j_x  = -\mu_x\\
&j_x \cos\varphi - j_z \sin\varphi = \mu_x \cos\varphi - \mu_z \sin\varphi
\end{align}
By eliminating $j_x$, $\mu_x$, $j_z$ and $\mu_z$, we obtain:
\begin{align}
\mupA-\jA &= \cos\varphi \left(\mupB - \jB\right)\\
\mupB+\jB &= \cos\varphi \left(\mupA + \jA\right)
\end{align}

The last set of equations that we need are the boundary conditions at the reservoirs. They read:
\begin{align}
    &j^c + \muc^L = eV_b\\
    &j^c - \muc^R = 0 \\
    &\jA + \mu_\parallel^L = 0 \\
    &\jB - \mu_\parallel^R = 0 \\
    &j^q + \tp^L = \k \Delta T \\
    &j^q - \tp^R = 0
\end{align}
with $\muc^{L/R}$, $\mu_\parallel^{L/R}$, $\tp^{L/R}$ the value of the potential, spin-resolved potential and temperature infinitely close to the left (L) and right (R) reservoir.

Finally, after some algebra, we can obtain the expressions of the currents and potentials:
\begin{widetext}
\begin{align}
j^c &= FG\left[ GYeV_b+S\k\Delta T \right]\\[5pt]
j^q &= FG\left[SeV_b+ K \k\Delta T\right]\\[5pt]
\jA &=F\gamma_A\left[GYeV_b+S\k\Delta T\right]+FK\delta_A\k\Delta T\\[5pt]
\jB &=F\gamma_B\left[GYeV_b+S\k\Delta T\right]+FK\delta_B\k\Delta T\\[5pt]
\muc &= 2r_B F\left(\left[(1+\X{B})G-\beta_B\gamma_B\right]\left[GYeV_b+S\k\Delta T\right] - \left[\sp_{B}G+\dsp_{B}\gamma_B\right]K\k\Delta T\right)\\[5pt]
\mupA &=F\left[\beta_A\left(1+\dfrac 1 {r_B}\right)-\beta_B\cos\angle-\gamma_A\right]\left[GYeV_b+S\k\Delta T\right] + F\left[\dsp_{A}\left(1+\dfrac 1 {r_B}\right)-\dsp_{B}\cos\angle-\delta_A\right]K\k\Delta T\\[5pt]
\mupB &=-F\left[\beta_B\left(1+\dfrac 1 {r_A}\right)-\beta_A\cos\angle-\gamma_B\right]\left[GYeV_b+S\k\Delta T\right] - F\left[\dsp_{B}\left(1+\dfrac 1 {r_A}\right)-\dsp_{A}\cos\angle-\delta_B\right]K\k\Delta T\\[5pt]
\tp &= 2r_B F\left[\left(\dfrac{1-\beta_B^2}{\L_B}+\X{B}\right)\left(SeV_b+K\k\Delta T\right)-Y\left(\sp_{B}G+\dsp_{B}\gamma_B\right)eV_b\right]
\end{align}
\end{widetext}
with the following notations:
\begin{align}
G &= \dfrac 1 2 \left(\sin^2 \angle + \dfrac 1 {r_A}+\dfrac 1 {r_B} +\dfrac 1 {r_A r_B}\right)\\
Y &=\left(\dfrac{1-\beta_B^2}{\L_B}+\X{B}\right)\X{A} + \left(\dfrac{1-\beta_A^2}{\L_A}+\X{A}\right)\X{B}\\
\gamma_i &= \left(\dfrac 1 2 \sin^2\angle + \X{j}\right)\beta_i + \X i \beta_j\cos\angle\\
\delta_i &= \left(\dfrac 1 2 \sin^2\angle + \X{j}\right)\dsp_{i} + \X i \dsp_{j}\cos\angle
\end{align}

with $(i,j)=(A,B)$ or $(B,A)$

\begin{align}
K &= \dfrac 1 2 \left(\dfrac 1 {r_A} + \dfrac 1 {r_B} +\dfrac 1 {r_A r_B}\right)    G - \dfrac{r_B\beta_B\gamma_B+r_A \beta_A\gamma_A}{2r_Ar_B}\\
S &= \X{A}\left(\sp_{B}G+\dsp_{B}\gamma_B\right) + \X{B}\left(\sp_{A}G+\dsp_{A}\gamma_A\right)\\
F &= \X{A}\X{B}\dfrac{1}{KGY}
\end{align}

The torque on layer B is defined by $\bv \tau=\bv J_N-\bv J_B = \dfrac{2\hbar}{e^2 \Rsh}\left(j_x \e x + j_z\e z - \bv j_{\parallel B} \right)=\dfrac{2\hbar}{e^2 \Rsh} \tau \bv{e}_1$, where $\bv{e}_1$ is the in-plane normal vector orthogonal to the magnetization of $F_B$. We obtain,

\begin{align}
    &\tau = -\dfrac 1 2 \sin \varphi \left(\mupA +\jA\right)\nonumber\\
    &=-\dfrac F 2 \sin\varphi\left\lbrace\left[\beta_A(1+\dfrac{1}{r_B})-\beta_B\cos\varphi \right] \left(GYeV_b+S\k\Delta T \right)\right.\nonumber\\
    & +\left.\left[\dfrac{\Delta s_{A}}{\L_A}(1-\beta_A^2)(1+\dfrac 1{r_B})-\dfrac{\Delta s_{B}}{\L_B}(1-\beta_B^2)\cos\varphi \right]K\k\Delta T\right\rbrace
    \label{eq:torque SM}
\end{align}

The expression of the waviness angle is, for any applied temperature gradient and/or voltage:
\begin{widetext}
\begin{align}
\cos \varphi_\ast = \dfrac{\beta_A(1+\dfrac 1{r_B})\left(GYeV_b+S\k\Delta T \right)+\dfrac{\Delta s_{A}}{\L_A}(1-\beta_A^2)(1+\dfrac 1{r_B})K\k\Delta T}{\beta_B\left(GYeV_b+S\k\Delta T \right)+\dfrac{\Delta s_{B}}{\L_B}(1-\beta_B^2)K\k\Delta T}
\end{align}
\end{widetext}
\section{Material parameters}
For the sake of completeness, we provide the parameters used for the numerical simulations. We used $\Rsh=2\ f\Omega\cdot \rm m^2$, and the values given in the following tables:
\setlength{\tabcolsep}{6pt}
\begin{table}[!h]
    \begin{tabular}{c|cccccc}
    Bulk & $\rho_\ast$ & $\beta$ & $\lsf$ & $s$ & $\ds$ & $\L$ \\
    material & [$\Omega\cdot$nm] & & [nm]& & & \\[5pt]
    \hline\\[-8pt]
    Cu &5&0&500&0.0185&0&$\pi^2/3$ \\[5pt]
    Co &75&0.46&60&-0.25&-0.02&$\pi^2/3$ \\[5pt]
    Py &291&0.76&5.5&-0.21&-0.044&$\pi^2/3$
    \end{tabular}
    \caption{material parameters for the bulk materials}
\end{table}

\begin{table}[!h]
    \begin{tabular}{c|cccccc}
    Interface & $r_b^\ast$ & $\gamma$ &$\delta$ &$\rm T_\mx$ &$\rm R_\mx$&$\L$  \\
    material& [$f\Omega\cdot \rm m^2$]  &     &   &   &   &    \\[5pt]
    \hline\\[-8pt]
    $\rm Cu|Co$ & 0.51 &0.77 & 0 & 0 & 0 & $\pi^2/3$    \\[5pt]
    $\rm Cu|Py$ & 0.5  &0.7  & 0 & 0 & 0 & $\pi^2/3$
    \end{tabular}
    \caption{material parameters for the interfaces}
\end{table}

\end{document}